\documentclass{iau} 
\usepackage{graphicx}

\title[Planetary Nebulae Population in the Virgo cluster ] 
{\large Planetary Nebulae and their parent stellar
  populations. Tracing the mass assembly of M87 and Intracluster light
  in the Virgo cluster core }

\author[Magda Arnaboldi \etal\ ] 
{Magda Arnaboldi$^{1,2}$, Alessia Longobardi$^3$, \and Ortwin Gerhard$^3$}

\affiliation{
  $^1$ESO, K. Schwarzschild Str. 2, 85748 Garching, Germany\\ email: {\tt marnabol@eso.org} \\[\affilskip]
  $^2$INAF, Oss. Astr. di Pino Torinese, 10025 Pino Torinese, Italy \\[\affilskip]
  $^3$Max-Planck-Institut f\"ur Extraterrestrische Physik, Postsach 1312, 85741 Garching, Germany}

\pubyear{2015}
\volume{317}  
\jname{The General Assembly of Galaxy Halos: Structure, Origin and Evolution}
\editors{A. Bragaglia, M. Arnaboldi, M. Rejkuba \& D. Romano, eds.}
\begin{document}

\maketitle

\begin{abstract}
The diffuse extended outer regions of galaxies are hard to study
because they are faint, with typical surface brightness of $1\%$ of
the dark night sky. We can tackle this problem by using resolved star
tracers which remain visible at large distances from the galaxy
centers. This article describes the use of Planetary Nebulae as
tracers and the calibration of their properties as indicators of the
star formation history, mean age and metallicity of the parent stars
in the Milky Way and Local Group galaxies . We then report on the
results from a deep, extended, planetary nebulae survey in a $0.5$
deg$^2$ region centered on the brightest cluster galaxy NGC 4486 (M87)
in the Virgo cluster core, carried out with SuprimeCam@Subaru and
FLAMES-GIRAFFE@VLT. Two PN populations are identified out to 150 kpc
distance from the center of M87. One population is associated with the
M87 halo and the second one with the intracluster light in the Virgo
cluster core. They have different line-of-sight velocity and spatial
distributions, as well as different planetary nebulae specific
frequencies and luminosity functions. The intracluster planetary
nebulae in the surveyed region correspond to a luminosity of four
times the luminosity of the Large Magellanic Cloud. The M87 halo
planetary nebulae trace an older, more metal-rich, parent stellar
population. A substructure detected in the projected phase-space of
the line-of-sight velocity vs. major axis distance for the M87 halo
planetary nebulae provides evidence for the recent accretion event of
a satellite galaxy with luminosity twice that of M33. The satellite
stars were tidally stripped about $1$~Gyr ago, and reached apocenter
at a major axis distance of $60-90$ kpc from the center of M87. The
M87 halo is still growing significantly at the distances where the
substructure is detected.

\keywords{Stars: AGB and post-AGB. (ISM) Planetary nebulae:
  general. Galaxies: general, abundances, Virgo cluster, elliptical
  and lenticular, cD, halos, formation, NGC~4486, kinematics and
  dynamics, structure. }
\end{abstract}

\firstsection 
\section{Introduction}
Accretion events are believed to be responsible for the build up of
stellar halos in elliptical galaxies (\cite[Delucia \& Blaizot
  2007]{DLB07}) at relatively low redshift ($z<2$; \cite[Oser
  \etal\ 2010]{O+10}). In the dynamical centers of galaxy clusters,
brightest cluster galaxies (BCGs) are expected to have the majority of
their stars accreted (\cite[Cooper \etal\ 2015]{Coo+15}). The galaxy
NGC~4486 (M87) is one of the nearest BCGs (at D=14.5 Mpc) in the
densest region of the Virgo cluster (\cite[Binggeli
  \etal\ 1987]{bing87}).  Its halo represents a benchmark for studies
of the assembly history of extended halos in high density
environments. Because of the large orbital time in their outer
regions, these halos may still contain fossil records of the mass
accretion events that lead to their hierarchical build up.

M87 has been the target of several imaging surveys and its close
proximity made it possible to identify planetary nebulae (PNs) with 8
meter class telescopes (\cite[Arnaboldi \etal\ 2003]{arna+03},
\cite[Aguerri \etal\ 2005]{Aguerri+05}, \cite[Castro-Rodriguez
  \etal\ 2009]{castro+09}). The goal is to use PN as kinematic tracers
and their general PN population properties as probes for star
formation history, age and metallicity of the parent stellar
population in those regions of M87 where the surface brightness is too
low to carry out absorption line spectroscopy.

In the next sections, we present a concise summary of the general
characteristics of PN population as tracers of stars and motions in
galaxies and then describe the results from the extended PN survey in
M87.

\section{General properties of PN population}\label{Sec2}
PNs are the final evolutionary stage for most
stars in the mass range $1-8\,\mbox{M}_\odot$. In the Milky Way (MW),
about 95\% of the stars will end their lives as PNs, while only 5\%
explode as supernovae. The PN phase lasts $\tau_{PN} \sim 3\times
10^4$ years at most, and its duration depends on the age and
metallicity of the parent stellar population (\cite[Buzzoni
  \etal\ 2006]{BAC2006}). $\tau_{PN}$ is also related to the expansion
time of a nebular shell $\tau_{PN} = D_{PN} / V_{exp}$, where $D_{PN}$
is the diameter and $V_{exp}$ is the expansion velocity of a PN shell;
typical expansion velocities for the brightest PNs are in the range
$11 - 22$ kms$^{-1}$ (\cite[Arnaboldi
  \etal\ 2008]{arna+08}). $\tau_{PN}$ could be shortened by the
presence of a hot interstellar medium (\cite[Dopita
  \etal\ 2000]{dopmar00}, \cite[Villaver \& Stanghellini
  2005]{Vilsta05}) which may remove the gaseous shell during its
expansion.

Because the diffuse nebula around the core is very efficient in
re-emitting $\sim 15\%$ of the UV energy radiated by the central star
in the optical Oxygen forbidden line [OIII] at $5007$ \AA\ [which is
  the brightest optical emission of a PN (\cite[Dopita
    \etal\ 1992]{Dopita+92})], PN stars can be efficiently selected
via narrow band imaging centered on the Oxygen line.

There are about 2000 PNs known out of 200 billion stars in the MW, and
they are mostly concentrated towards the MW plane. A typical Galactic
PN has an average shell diameter of about $0.3$ pc. Hence when a
sample of PNs similar to those in the MW are detected in external
galaxies at distances larger than $1$ Mpc, they are identified as
spatially unresolved emissions of monochromatic green light at
5007\AA.

The integrated [OIII] flux $F_{5007}$ of a spatially unresolved PN can
be expressed as $m_{5007}$ magnitude via the formula:
\begin{equation}
m_{5007} = -2.5 \log(F_{5007}) - 13.74 \label{m5007}
\end{equation} 
(\cite[Jacoby 1989]{Jac89}). Narrow band imaging of external
galaxies provide $m_{5007}$ magnitudes for the entire PN population of
the surveyed galaxy, down to a given limiting flux. It is then
possible to derive the Planetary Nebulae luminosity function (PNLF)
$N(m_{5007})$ for that PN population at the galaxy distance. The PNLF
has been used extensively as secondary distance indicator in early and
late-type galaxies within $10-15$ Mpc distance (see \cite[Ciardullo
  \etal\ (2002)]{Ciardullo+2002} for a review). The PNLF is often
approximated by an analytical formula given by
\begin{equation}
N(M) \propto e^{{0.307}M}\times \left(1-e^{3(M^* - M)}\right) \label{PNLF}
\end{equation}
as introduced by \cite[Ciardullo \etal\ (1989)]{Ciardullo+89}, where
$M^* = -4.51$ is the absolute magnitude of the bright cut-off of the
PNLF (\cite[Ciardullo \etal\ 1998]{Ciardullo+98}). This analytical
formula is the product of two exponential terms: the first term can be
thought of as the dimming of the [OIII] flux as the shell expands at
uniform speed (\cite[Heinze \& Westerlund 1963]{HenWest1963}), and the
second term models the cut-off at bright magnitudes.  Hence the
formula in Eq.~\ref{PNLF} describes a PN population as an ensemble of
diffuse expanding shells powered by unevolving massive cores, all at
about $M_{core} \simeq 0.7 M_\odot$, which are emitting a total
luminosity of $L\simeq 6000 \mbox L_\odot$ (\cite[Ciardullo
  \etal\ 2002]{Ciardullo+2002}).

Simple stellar population theory predicts that PN cores should become
fainter as the stellar population ages, with core masses as low as
$M_{core} \leq 0.55 M_\odot$ in a 10 Gyr old, solar-enriched, stellar
population (\cite[Buzzoni \etal\ 2006]{BAC2006}). \cite[Marigo
  \etal\ (2004)]{Marigo+2004} computed a relative dimming of 4
magnitudes for the bright cut-off $M^*$ of the PNLF for a 10 Gyr old
population with respect to that of a 1 Gyr old population.  Such a
strong dependence of $M^*$ on age is {\bf not} observed though:
empirically, the absolute magnitude of the PNLF bright cut-off is the
same for PN populations in star forming disks and in ellipticals
(\cite[Longobardi \etal\ (2013)]{L+13}). A possible explanation is that
binary stars are progenitors to the brightest PNs in old populations
(\cite[Ciardullo \etal\ 2005]{C+05}). Still there may be systematic
variations of the PNLF that correlate with the star formation history,
mean age and metallicity of the parent stellar population, which can be
used to constrain their values and any spatial variations in extended
stellar halos.

\subsection{PN visibility lifetime and luminosity functions in the Milky Way and Local Group galaxies}

\underline{PN specific frequency} - The total number of PNs associated
with the bolometric luminosity of a parent stellar population is
expressed as $N_{PN} = \alpha L_{\odot,bol}$, where $\alpha$ is the PN
specific frequency. The value of the $\alpha$ parameter is related to
the normalization in Eq.~\ref{PNLF}; i.e. to the observed total number of
PNs of a given detected population. The value of $\alpha$ is related
to the PN visibility lifetime $\tau_{PN}$ by the equation
\begin{equation}
\alpha = \frac{N_{PN}}{L_{\odot,bol}} = {\rm B} \tau_{PN} \label{TPN}
\end{equation}
where ${\rm B}$ (independent of the stellar population) is the PN
formation rate (stars/yr/$L_\odot$; \cite[Buzzoni
  \etal\ 2006]{BAC2006}).  The measured values for the $\alpha$
parameter show strong scatter for stellar populations redder than
$(B-V) \ge 0.8$ (\cite[Coccato \etal\ 2009]{Coccato+09}, \cite[Cortesi
  \etal\ 2013]{Cortesi+13}) with an inverse correlation with the far
ultraviolet (FUV) color excess, that is stellar populations with a
strong FUV excess are {\it PN-starved}. Stellar populations with an UV
up-turn or FUV color excess are the old and metal rich populations
in massive elliptical galaxies. Differently, stellar populations in
irregular galaxies like the LMC are {\it PN-rich}. In general PN
populations show systematic variations of the $\alpha$ values with the
integrated photometric properties of the parent stellar population,
hence variations of the measured $\alpha$ values as function of
radius in an extended stellar halo can be used as a signal for
different stellar populations in the halo.

\underline{PNLF morphology} - The PNLF shows systematic variations
that correlate with the average age and metallicity of the parent
stellar population. In the $\log(N)$ vs. $m_{5007}$ plot, the gradient
of the PNLF within $2.5$ magnitude below the brightest can be steeper
or shallower than $0.307$ (as in Eq.~\ref{PNLF}). The PN population
associated with the MW bulge has a steeper PNLF than that derived for
the M31 PNs (Arnaboldi et al. 2015, in prep.), while those in the outer
regions of star-forming disks have a shallower or no gradient
(\cite[Ciardullo \etal\ 2004]{ciard+04}).  \cite[Longobardi
  \etal\ 2013]{L+13} proposed a generalized formula for the PNLF
according to the equation
\begin{equation}
N(M)=c_1 e^{c_2M}\times \left(1-e^{3(M^* - M)}\right); M^*=-4.51
\label{genPNLF}
\end{equation}
where $c_1$ is related to the value of the $\alpha$ parameter to first
order and $c_2$ is related to the gradient of the $\log(N)$ at faint
magnitudes. We are planning to apply the generalized formula to
complete and extended PN populations and to correlate the $c_1,\,c_2$
derived values with the mean ages and metallicities of the parent
stellar populations from spectroscopic measurements, so that we can
have a better calibration of the PN probes.

In addition to the gradient, the PNLF of stellar population in low
luminosity - metal poor galaxies - shows the presence of a {\it
  dip}. This feature of the PNLF is measured with high significance in
the well sampled PNLF for the LMC (\cite[Reid \& Parker
  2010]{LMC2010}), the SMC (\cite[Jacoby \& De Marco (2002)]{SMC02}),
NGC~6822 (\cite[Hern\'andez-Mart\'inez \& Pe\~na 2009]{NGC6822}) and
in the outer regions of M33 (\cite[Ciardullo
  \etal\ 2004]{ciard+04}). This dip falls within an interval of $2$ to
$4$ magnitudes below $M^*$, but the magnitude at which the dip is
detected varies between the PN populations sampled in the Local Group.

\underline{Strategy} - we can use the global properties of the PN
populations, their PNLFs, the gradients, dips and $\alpha$ values to
signal transition from old/metal-rich to fading/metal-poor populations
when the individual stars cannot be resolved or their surface
brightness is too low to carry out integrated light photometry or
absorption line spectroscopy.

\section{The PN populations in the Virgo cluster core}

In 2010 we started an imaging survey with SuprimeCam@Subaru to cover
$0.5$ deg$^2$ in the M87 halo; at the distance of the Virgo cluster
(D=14.5 Mpc) this is equivalent to an area of ($130$ kpc)$^2$. We
wanted to use the general properties described in Sec.~\ref{Sec2} to
study its stellar population and kinematics.

We acquired deep narrow band images centered on the [OIII] emission
redshifted to the systemic velocity of M87 ($V_{sys} = 1275$
kms$^{-1}$) and deep off-band images in the V-band. We identified PN
candidates as spatially unresolved [OIII] sources with no continuum
following the procedure described in \cite[Arnaboldi
  \etal\ (2002)]{arna+02}. The final magnitude limited catalog consisted
of $688$ PN candidates down to 2.5 magnitude from the brightest PN
(\cite[Longobardi \etal\ 2013]{L+13}). We then carried out the
spectroscopic follow-up with FLAMES/GIRAFFE@VLT to acquire spectra
for these candidates. We obtained spectra for $289$ confirmed PNs
(\cite[Longobardi \etal\ 2015a]{L+15a}) which we analyzed together
with $12$ previous identified PNs from \cite[Doherty
  \etal\ (2009)]{D+09}, for a total combined sample of $301$ PNs.

\begin{figure}
\begin{center}
 \includegraphics[width=3.0in]{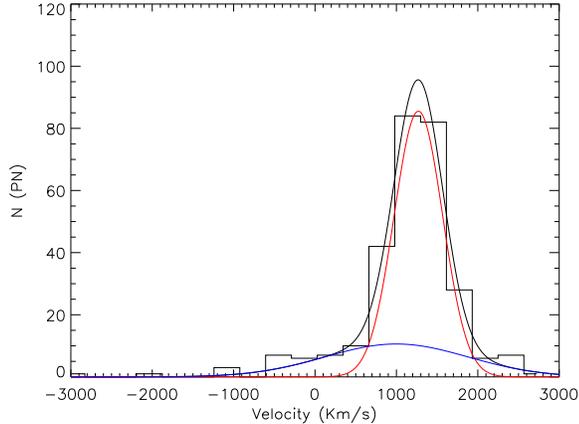}
\vspace*{-.25 cm}
 \caption{Histogram of the line-of-sight velocities of the
   spectroscopically confirmed PNs (black histogram) fitted with a
   double Gaussian (black curve). Red and blue lines represent the two
   Gaussian associated with the M 87 halo and the IC components. From
   \cite[Longobardi \etal\ (2015a)]{L+15a}.
   \label{fig1}}
\end{center}
\end{figure}

\begin{figure}
\begin{center}
 \includegraphics[width=3.0in]{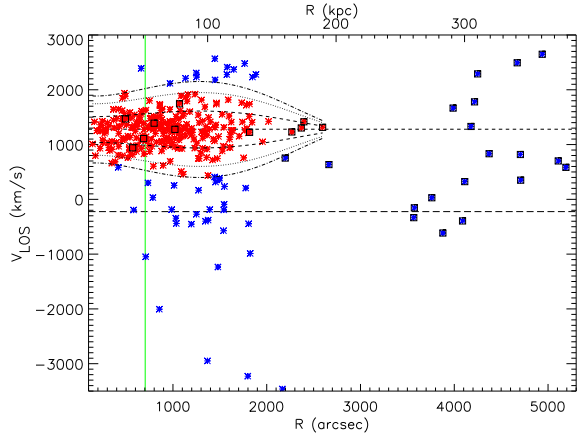}
\vspace*{-.25 cm}
 \caption{Projected phase-space diagram, $V_{LOS}$ vs. major axis
   distance from the center of M 87, for all spectroscopically
   confirmed PNs in the Virgo core.The major axis distance is given
   both in arcsec (bottom axis), and in kpc (top axis), where 73 pc$ =
   1''$. The PNs are classified as M 87 halo PNs (red asterisks) and
   ICPNs (blue asterisks), respectively. Black squares identify
   spectroscopically confirmed PNs from Doherty et al. (2009). The
   dashed horizontal line shows the M 87 systemic velocity $V_{sys} =
   1275$ kms$^{-1}$, while the continuous green line shows the
   effective radius $R_e = 703.914''$ (\cite[Kormendy
     \etal\ 2009]{K+09}). From \cite[Longobardi \etal\ (2015a)]{L+15a}.
   \label{fig2}}
\end{center}
\end{figure}

We built the line-of-sight velocity distribution (LOSVD) for the whole
PN sample: this is shown in Fig.\,\ref{fig1}. The LOSVD is
characterized by a strong peak at $1275$ kms$^{-1}$ and a
second moment $\sigma_n \simeq 300$ kms$^{-1}$, with large asymmetric
wings, with a tail extended towards zero and negative LOS
velocities. These wings can be modeled with a broad Gaussian
component, centered at $v_b=995$ kms$^{-1}$ and $\sigma_b=900$
kms$^{-1}$.
The narrow component is consistent with the LOSVD for the stars in the
M87 halo, while the broad component has a LOSVD similar to that of
Virgo cluster galaxies. We thus selected the M87 halo PNs as those
associated with the narrow component in the LOSVD by means of a robust
iterative procedure (\cite[Longobardi \etal\ 2015a]{L+15a}). The PNs
in the broad asymmetric wings of the LOSVD are thereby tagged as
Virgo intracluster (IC) PNs. In Figure~\ref{fig2}
we show the projected phase space diagram for the 301 PNs around M87 and the
additional ICPNs from \cite[Doherty \etal\ (2009)]{D+09}.

We studied the spatial distribution of the two selected PN samples and
compared their profiles with the V band photometry of M87 by
\cite[Kormendy \etal\ (2009)]{K+09}. The number density profile of the
M87 halo PNs follows the Sersic $n=11.8$ surface brightness profile,
while that of the ICPNs follows a much flatter radial profile, which
is consistent with a power law, $N_{ICPN}(R) \propto R^\gamma $ with
$\gamma = [-0.34,-0.04]$. By scaling the M87 PN density profile to the
inner regions we derived the value of the $\alpha$ parameter for the
M87 PN halo population: $\alpha_{halo}= (1.06\pm0.12) \times 10^{-8}
PN L^{-1}_{\odot,bol}$. The value of the $\alpha$ parameter for the
intracluster light is different: the same procedure applied to the
ICPN number density profile returns $\alpha_{ICL}= (2.72\pm 0.72)
\times 10^{-8} PN L^{-1}_{\odot,bol}$.

We then compared our $\alpha_{halo}$ and $\alpha_{ICL}$ values with
those for PN populations in nearby galaxies. Galaxies with $(B-V) \leq
0.8$ have PN specific frequency values similar to that of
$\alpha_{ICL}$.  For redder galaxies, the scatter increases; the
value of $\alpha_{halo}$ is the same as for populations with an
FUV excess that are PN-starved. Hence the IC component contributes
three times more PN per unit bolometric luminosity than the M87 halo
light, signaling a change of population from halo to ICL. This
transition is consistent with the existence of a color gradient
towards bluer colors in M87 at large radii (\cite[Rudick
  \etal\ 2010]{R+10}), and an ICL population that is mostly old (age
$\simeq 10$ Gyr) with a mean metallicity of $[M/H] \simeq -1.0$
(\cite[Williams \etal\ 2007]{wills+07}).
On the basis of the kinematic separation of the M87 halo and
intracluster PNs, we built the two independent PNLFs, and examined
them in turn. The PNLF for the M87 halo PNs is steeper than in
Eq.~\ref{PNLF}. We used the generalized formula from Eq.~\ref{genPNLF}
and derived $c_2 = 0.72$ and a distance modulus of $m-M = 30.8$.  The
PNLF of the ICPNs shows a dip at $1-1.5$ mag below $M^*$, reminiscent
of the morphology of the PNLF observed in PN populations associated
with low luminosity/metal poor galaxies. We fitted the generalized
PNLF formula to the PNLF of the ICPNs, and found $c_2 = 0.66$ and a
distance modulus of $m-M = 30.8$.

We could estimate the luminosity associated with the ICPNs by
integrating the ICL surface brightness profile over the surveyed area,
which amounts to $L_{ICL}= 5.3 \times 10^9 L_{\odot,bol}$. From the
presence of the ``dip'' of the PNLF, the parent stellar population is
similar to that of the Magellanic Clouds or in the outer regions of
the M33 disk. Thus the sampled luminosity of the ICL is equivalent to
four time the luminosity of the $LMC$ or 1.5 times that of $M33$, over
the whole surveyed area of ($130$ kpc)$^2$.

\section{The late mass assembly of the M87 halo}

We then turned to the M87 halo PNs and studied the projected
phase-space diagram $v_{LOS}\, \mbox{vs.}\, R_{Maj}$ for this
component. The density of points is not uniform in the range of
velocities covered by the M87 halo. There is a clearly identifiable
{\bf V}-shaped over-density, or {\it ``Chevron''}, between 30 and 90
kpc radii, with its vertex, or edge, culminating at $90$ kpc or
$1200''$ (\cite[Longobardi \etal\ 2015b]{L+15b}) .  These high density
substructures in phase-space are likely to be associated with the
disruption of a satellite galaxy in the deeper potential of a massive
host (\cite[Quinn 1984]{QJ1984}).

\begin{figure}
\begin{center}
     \includegraphics[width=8.5cm, clip=true, trim=0.9cm 0cm 0.5cm 0.cm]{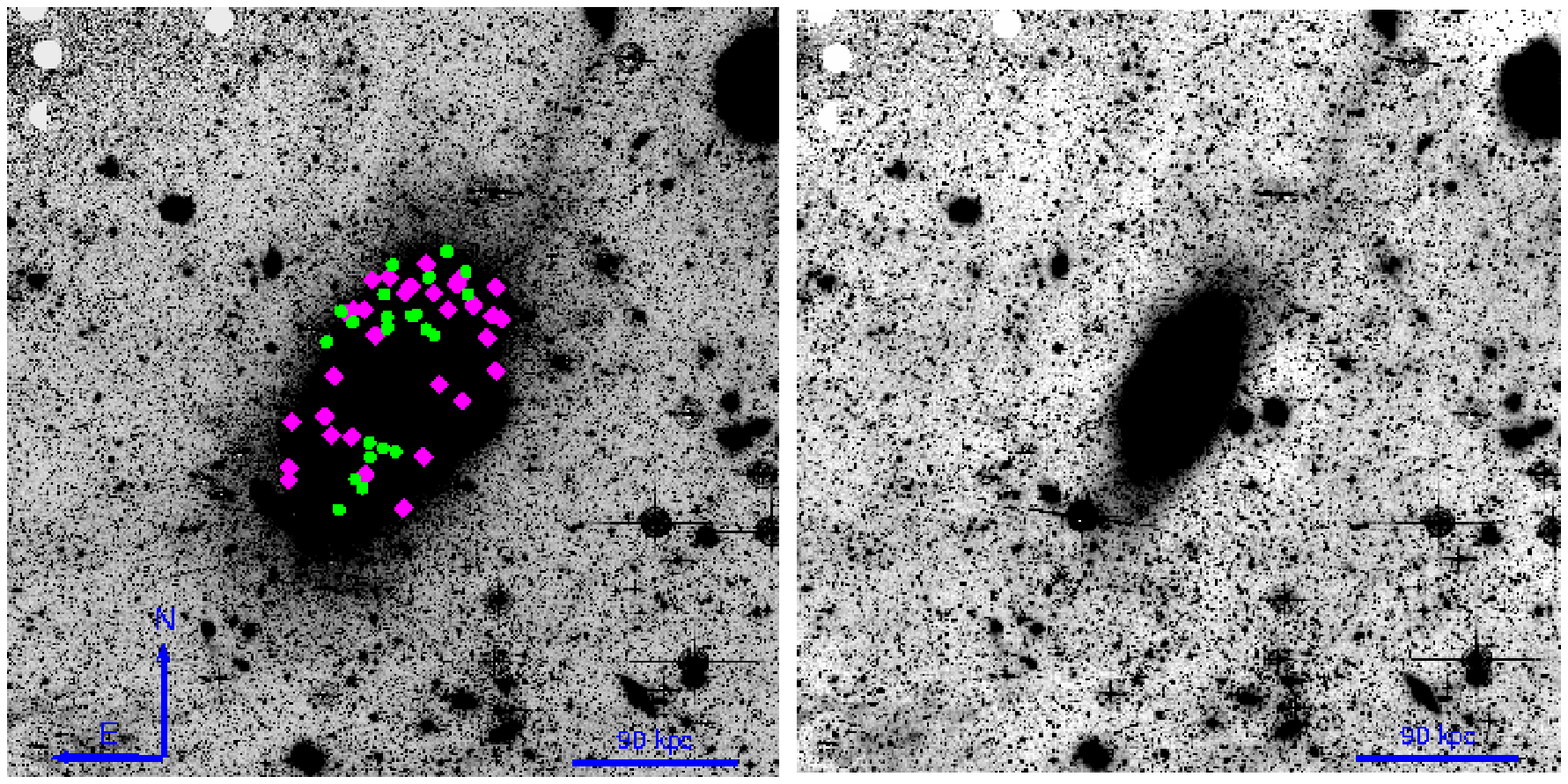}
\hspace{-0.1cm}
    \includegraphics[width=4.2cm, clip=true, trim=0cm 0.cm 0.2cm 0.cm]{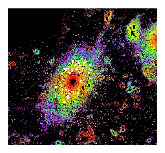}
\vspace*{-.20 cm}
 \caption{ Spatial and color distribution associated with the
   kinematic substructure identified in the phase space of the M 87
   halo PNs. Left panel: V-band image of a $1.6\times1.6$ deg$^2$
   centered on M 87 from Mihos et al. (in prep.). Full circles, and
   diamonds indicate the spatial positions of the M 87 halo PNs in the
   chevron substructure. Magenta and green colors indicate PN LOSVs
   above and below $V_{LOS} = 1254$ kms$^{-1}$, the LOSV at the end of
   the Chevron. Central panel: unsharp masked image of M 87 median
   binned to enhance faint structures. The crown-shaped substructure
   is visible at distance of $800''-1200''$(60−90 kpc) along the major
   axis, NW of M 87. The blue line measures 90 kpc. Right panel: (B −
   V) color image of M 87 from Mihos et al. (in prep.) with chevron
   PNs overplotted (white dots). The dashed ellipse indicates the
   isophote at a major axis distance of $1200''$. The crown is found
   in a region where the (B − V) color is on average 0.8, bluer than
   on the minor axis.  From \cite[Longobardi \etal\ (2015b)]{L+15b}.
   \label{fig3}}
\end{center}
\end{figure}

We looked at the LOSVD in three radial bins over the radial range of
the Chevron, and employed a Gaussian Mixture Model to assign each M87
halo PN a probability to belong to the {\bf V} substructure or to the
smooth halo. A total of 54 PNs are thereby associated to the {\bf V}
substructure and 200 PNs to the smooth halo component
(\cite[Longobardi \etal\ 2015b]{L+15b}). We then looked at the spatial
distribution of the Chevron PNs: the highest density occurs at
$1200''$ NW along the M87 major axis, and is spatially correlated with
a diffuse, extended substructure, labeled ``the crown'' of M87
(\cite[Longobardi \etal\ 2015b]{L+15b}), see Figure~\ref{fig3}. The
luminosity of the substructure is about $60\%$ of the light at the
location where the crown is found. We computed the total luminosity of
the disrupted satellite from the 54 PNs: it amounts to $L_{sat} = 2.8
\times 10^9 L_{\odot,bol}$. We also looked at correlation between the
high density of the chevron PNs and the (B-V) color in the M87 halo. A
strong correlation is found, with bluer color (B-V = 0.76) measured at
the position of the ``crown''. From the luminosity and the color we
infer that the luminosity of the dissolved satellite is equivalent to
about twice the luminosity of M33.

From the distribution and velocities of Chevron PNs in
Figure~\ref{fig3}, a possible interpretation of the satellite orbit
could be that it was first disrupted entering M87 from the south
(along the green dots), with the debris then moving up north, turning
around in the crown region, coming back down on both sides across M87
(magenta dots). The velocities would then imply that the northern
side of M87 is closer to the observer. The dynamical time for such an
orbit is of the order of 1 Gyr (\cite[Weil \etal\ 1997]{W+1997}).

\section{Conclusions}

Using PNs as tracers we showed that the stellar halo of the BCG M87 is
distinct from the surrounding ICL in its kinematics, density profile,
and parent stellar population, consistent with the halo of M87 being
redder and more metal-rich that the ICL. The ICL in our surveyed
fields corresponds to about four times the luminosity of the LMC,
spread out over a region of (130 kpc)$^2$. It is remarkable that
population properties can be observed for such a diffuse
component. Based on its population properties we propose that the
progenitors of the Virgo ICL were low-mass star forming galaxies.

We also presented kinematic and photometric evidence for an accretion
event in the halo of M87. This event is traced by PNs whose phase
space shows a distinct chevron-like feature, which is the result of
the incomplete phase-space mixing of a disrupted galaxy. At a major
axis distance of $R\sim 69-90$ kpc where the width of the chevron goes
to zero, a deep optical image shows the presence of a crown-like
substructure that contribute more than $60\%$ of the light in this
area. The luminosity of the satellite corresponds to about two times
M33 with color $(B-V) = 0.76$. The similar colors of the accreted
satellite and ICL suggest that the halo of M87 is presently growing
through the accretion of similar star-forming systems as those that
build up the diffuse ICL component. The newly discovered substructure
within the halo of M87 demonstrates that beyond a distance of 60 kpc
its halo is still assembling.

\section{Acknowledgments}
The authors wish to thank J.C Mihos, R. Hanuschik for their contribution,
and the time allocation committees of the Subaru Telescope and the ESO
OPC for the opportunity to carry out this exciting project. Based on
observations made with the VLT at Paranal Observatory under programmes
088.B-0288(A) and 093.B-066(A), and with the Subaru Telescope under
programme S10A-039.

\end{document}